# The $(11\bar{2}2)$ and $(\bar{1}2\bar{1}6)$ twinning modes modelled by obliquity correction of a (58°, a+2b) prototype stretch twin

Cyril Cayron

Laboratory of ThermoMechanical Metallurgy (LMTM), PX Group Chair, Ecole Polytechnique Fédérale de Lausanne (EPFL), Rue de la Maladière 71b, 2000 Neuchâtel, Switzerland

**Abstract**

The $\{11\bar{2}2\}$ and $\{11\bar{2}6\}$ twinning modes were recently put in evidence by Ostapovets *et al.* (Phil. Mag, 2017) and interpreted as $\{10\bar{1}2\} - \{10\bar{1}2\}$ double-twins formed by a simultaneous action of two twinning shears. We propose another interpretation in which the twinning modes result from a one-step mechanism based on the same (58°, **a**+**2b**) prototype stretch twin. The two twins differ from the prototype twin by their obliquity correction. The results are compared with the classical theory of twinning and with Westlake-Rosenbaum's model of $\{11\bar{2}2\}$ twinning. An unconventional twinning mode recently discovered in a magnesium single crystal based on the same prototype twin will be the subject of a separate publication.

# 1. Introduction

The $\{11\bar{2}2\}$ twinning mode is classical in titanium and zirconium alloys. The disorientation between the parent grains and the $\{11\bar{2}2\}$ twins is a rotation of 64° around a **a**+**2b** axis, simply noted here (64°, **a**+**2b**). The disorientations are used in Electron Back Scatter Diffraction (EBSD) to identify and quantify the twins. The $\{11\bar{2}2\}$ twins and the $\{1\bar{1}02\}$ extension twins are easily distinguished in EBSD; the former form (64°, **a**+**2b**) boundaries and the latter (86°, **a**) grain boundaries. The histograms of disorientations at grain boundaries extracted from the experimental EBSD data in titanium and zirconium alloys show that $\{11\bar{2}2\}$ and $\{1\bar{1}02\}$ twins are the most observed twinning modes, and that the frequency of $\{11\bar{2}2\}$ twins is larger (sometimes twice) than that of the $\{1\bar{1}02\}$ twins [1][2]. The $\{11\bar{2}2\}$ twins have been also studied by Transmission Electron Microscopy (TEM) [3][4]; the twins are unambiguously identified by the mirror symmetry across the $\{11\bar{2}2\}$ interface plane in the selected area diffraction pattern.

The $\{11\bar{2}2\}$ twins are usually considered to result from a shear along a $\{11\bar{2}2\}$ plane in the $\langle 11\bar{2}3 \rangle$ direction. The shear amplitude *s* depends on the ratio of lattice parameters $\gamma = \frac{c}{a}$; the formula reported in literature is

$$s = \frac{2}{3}\left(\frac{\gamma^2 - 2}{\gamma}\right) \qquad (1)$$

The value *s* = 0.219 is obtained by formula (1) with $\gamma = 1.588$. The origin of formula (1) is not always mentioned in the recent papers and should be traced back. Crocker and Bevis [5] showed that the general theory [6][7] leads to a shear amplitude of 0.958, which was considered too high. A smaller numerical value $s$ = 0.219 was then given by these authors in the same work [5] considering a supercell in which only one-third of the atoms come to their correct positions by the lattice shear while the other atoms in the cell shuffle (i.e. move independently of the shear). It seems legitimate to wonder why the $\{11\bar{2}2\}$ twins can be more frequent than $\{1\bar{1}02\}$ extension twins if the shear value of the former ($s \approx 0.219$) is larger than that of the latter ($s \approx 0.174$). This question could not be raised at that time because the EBSD statistics were not yet established. Formula (1) is the reference for $\{11\bar{2}2\}$ twinning reported in many classical review papers such as those of Crocker and Bevis [5], Christian and Mahajan [8], and Yoo [9], but these works don't specify where it comes from. Actually, formula (1) should be attributed to Westlake [10]. Westlake shows succinctly that the twin could be formed by sequence of dislocations on every third planes. Rosembaum took this model and added many more details on the atomic movements and dislocations [11]. Among the thirty three pages of Ref. [11], twenty one were devoted to the model of $\{11\bar{2}2\}$ twinning. This refined model, called in the rest of the paper Westlake-Rosenbaum (WR) model, is fundamentally based on the belief that the twinned structure should be built from the parent crystal by the coordinated gliding of partial dislocations on the $\{11\bar{2}2\}$ plane. These dislocations (called zonal dislocations) were inspired by the $\{11\bar{2}2\}\langle\bar{1}\bar{1}23\rangle$ slips observed a few years before in zinc and cadmium. Trying to find the mechanism of twinning and more generally of structural phase transformations by considering the so-called "twinning dislocations" was an approach that has continued to be developed with a renew interest that came from the introduction of the concept of "disconnection" twenty years ago [12], helped by the extensive use of molecular dynamic simulations [13]. Those developments clearly deviate from the initial concepts of displacive transformations in which all the atoms move collectively (militarily). Indeed, the primary theory of martensitic transformations [14][15] and deformation twinning [6][7] only use linear distortions (based on shear matrices). This situation is well summarized in the introduction of Ref. [16] "*there are two competing theories: (1) the classical theory in which a homogeneous shear and atomic shuffling have to be involved for twinning in HCP metals [...] (2) the interface disconnection model in which a multiplicity of 'twinning dislocations' are defined at twin boundaries*". In our recent works [17]-[20], we have followed the former classical theory in the same spirit as its pioneers [5]-[7]; we have just replaced the simple shear matrices by a more general concept of "angular-distortive matrices" in order to calculate the continuous paths of the atomic displacements and lattice distortion. We made no use of "transformation dislocations" or "twinning dislocations". The reasons of this choice are based on our conviction that these dislocations are the *consequences* of the distortion mechanism and not their *causes*; they are the defects let by transformation inside the surrounding matrix. Our point of view is detailed in Ref.[21].

The present paper is thus a logic continuation of our previous works on displacive transformations and deformation twinning. It seeks an alternative model to the $\{11\bar{2}2\}$ twinning which is not based on "twinning dislocations". Dislocations can be introduced in further works in order to understand how the surrounding matrix accommodates the twinning distortion, but this task is not the purpose of the present work. There is a second motivation for the paper. The $\{11\bar{2}2\}$ twins, so very frequently observed in titanium and zirconium alloys, are totally absent in magnesium alloys. Nonetheless, an interesting feature appears in some EBSD papers on magnesium AZ31 alloys: one can notice in the histograms of disorientations between neighbour grains that a peak is often present at angles of angle 58° (Fig 3 of Ref. [22]) or 56° (Fig.4b,c of Ref. [23]), with an rotation vector of type **a**+2**b**. Until recently, no interpretation could be given for this peak because it doesn't correspond to any theoretical twin modes predicted with classical twinning theory or disconnection theory. One may however wonder about a possible link between the (64°, a+2b) rotation characterizing the $\{11\bar{2}2\}$ twins in the Ti and Zr alloys and this (58°, **a**+2**b**) rotation observed in magnesium alloys. Does the peak at (58°, **a**+2**b**) correspond to a real twinning mode? Is this mode linked to the (64°, **a**+2**b**) twins? Ostapovets *et al.* [24] very recently published an interesting theoretical paper that establishes such a link. They observed by EBSD and TEM twins in polycrystalline rolled magnesium with $\{11\bar{2}2\}$ or $\{11\bar{2}6\}$ habit planes. They show that these twins are conjugate and they interpreted them as a $\{10\bar{1}2\} - \{10\bar{1}2\}$ double-twin formed by the simultaneous action of two twinning shears.

We came to the subject of the (58°, **a**+2**b**) twins during the summer 2016 because we wanted to explain some odd twins mapped by EBSD in a magnesium single crystal. We did not publish it at that time our theoretical work because the habit planes we observed were incompatible with either $\{11\bar{2}2\}$ or $\{11\bar{2}6\}$ planes. The habit planes of these odd twins are actually unconventional and associated with a derived form of the (58°, **a**+2**b**) twins. The experimental and theoretical study related to this unconventional twin will the subject of a separated publication [25]. The aim of the present paper is to explain the theoretical model we developed for the conventional $\{11\bar{2}2\}$ or $\{11\bar{2}6\}$ twins. This model will be also used in Ref. [25]. As Ostapovets *et al.* [24] our theoretical analysis lead us to establish a link between these twinning modes, but our work differs from that of these authors in the mechanism. Instead of introducing a double-twinning mechanism, we imagined that the twins could be obtained in one-step, following the same approach as we used for martensitic transformations and for extension and contraction twinning. We have shown that it is possible to model the (86°, **a**) extension twins [19] and the (56°, **a**) contraction twins [20] by using a prototype stretch distortion and by combining it with an additional and small (few degrees) rotation **R** called obliquity in order to make a plane untilted and restored when the twinning process is complete. This plane is the shear plane of the distortion matrix of complete transformation; it is also the habit plane for conventional twinning. The same method will be used here. However, contrarily the previous works [19][20], the distortion matrices will be calculated only between the initial and final states, but not during continuous process. Indeed, the algebraic equations of the atomic trajectories could be analytically determined only for some atoms, but not yet for all the atoms because of the high complexity of the calculations. The hard-sphere assumption applied in Refs. [19][20] is not used in the present paper. The (58°, **a**+2**b**) orientation will be used as a special "prototype" configuration in which the distortion matrix is triangular. It will appear that $\{11\bar{2}2\}$ and $\{11\bar{2}6\}$ twinning modes are conjugate modes derived from this prototype twin by obliquity correction. It will then be shown that the (64°, **a**+2**b**) twins result from a shear on a plane $\{11\bar{2}2\}$ along the direction $\langle\bar{1}\bar{1}23\rangle$ with an amplitude of *s* = 0.102, which is less than half the amplitude resulting from the WR model. This will explain why the $\{11\bar{2}2\}$ twins are more frequent than the $\{1\bar{1}02\}$ extension twins (*s* = 0.174). More importantly, the direction found in the new model is opposite to the WR's one. Possibilities to compare experimentally the new model and the WR's model will be proposed.

## 2. Crystallographic model of the (58°, a+2b) stretch twin

The ratio of lattice parameter is $\gamma = \sqrt{8/3} \approx 1.633$ for ideal hard-sphere packing. The values chosen for magnesium and titanium are $\gamma = 1.625$ and $\gamma = 1.587$, respectively. We call $\mathbf{B}_{hex} = (\mathbf{a}, \mathbf{b}, \mathbf{c})$ the usual hexagonal basis, and $\mathbf{B}_{ortho} = (\mathbf{x}, \mathbf{y}, \mathbf{z})$ the orthonormal basis represented in Fig. 1 and linked to $\mathbf{B}_{hex}$ by the coordinate transformation matrix $\mathbf{H}_{hex}$:

$$\mathbf{H}_{hex} = [\mathbf{B}_{ortho} \rightarrow \mathbf{B}_{hex}] = \begin{pmatrix} 1 & -1/2 & 0 \\ 0 & \sqrt{3}/2 & 0 \\ 0 & 0 & \gamma \end{pmatrix} \quad (1)$$

The matrix $\mathbf{H}_{hex}$ is commonly called "structure tensor" in crystallography. It can be used to express directions into the orthonormal basis $\mathbf{B}_{ortho}$. For planes, it is $\mathbf{H}^*_{hex}$, the inverse of its transpose, that should be used. We note O, the "zero" position that is let invariant by the distortion and we note X, Y and Z the atomic positions defined by the vectors **OX** = $\mathbf{a}$ = [100]$_{hex}$, **OY** = $\mathbf{a}$ + 2$\mathbf{b}$ = [120]$_{hex}$ and **OZ** = $\mathbf{c}$ = [001]$_{hex}$. It can be checked with the matrix $\mathbf{H}_{hex}$ that **OX** = [100]$_{ortho}$, **OY** = [0 $\sqrt{3}$ 0]$_{ortho}$ and **OZ** = [0 0 $\gamma$]$_{ortho}$. Other atomic positions are labelled according to the same notations used in our previous works [19][20]. The three-index notation will be preferred to the four-index notation because of the 3x3 matrix calculations. Only conventional plane will be noted in four-index in order to help the comparison with results published in literature.

Three important matrices define the crystallographic characteristics of a twin: the coordinate transformation matrix, the distortion matrix, and the correspondence matrix. The coordinate transformation matrix $\mathbf{T}^{p \rightarrow t}_{hex}$ is given by the coordinates of the reference basis of the twin related to the parent basis. The distortion matrix $\mathbf{D}^{p \rightarrow t}_{hex}$ is given by the coordinates of vectors of the parent basis after distortion, and expressed in the initial parent basis before distortion. The correspondence matrix $\mathbf{C}^{t \rightarrow p}_{hex}$ is made of the same vectors but now expressed in the reference basis of the twin. The three matrices are linked by a master equation:

$$\mathbf{C}^{t \rightarrow p}_{hex} = \left(\mathbf{T}^{p \rightarrow t}_{hex}\right)^{-1} \mathbf{D}^{p \rightarrow t}_{hex} \quad (1)$$

The correspondence matrix specifies in which direction of the twin crystal a direction of the parent crystal is transformed. This matrix is independent of any obliquity correction. This can be checked by the fact that $\mathbf{C}^{t \rightarrow p}_{hex}$ is invariant if $\mathbf{D}^{p \rightarrow t}_{hex}$ and $\mathbf{T}^{p \rightarrow t}_{hex}$ are changed into $\mathbf{R}.\mathbf{D}^{p \rightarrow t}_{hex}$ and $\mathbf{R}.\mathbf{T}^{p \rightarrow t}_{hex}$ with $\mathbf{R}$ a rotation matrix. We now detail the model and the associated calculations of the three crystallographic matrices. The analytical calculations were performed with Mathematica. The program is available in Supplementary Material *SupplData1*.

As shown in Fig. 1, the (58°, **a**+2**b**) configuration reported in literature in some disorientation histograms (see introduction) is special because:

- The parent vector $\mathbf{OX}_2 = [200]_{hex\_p}$ is parallel to the twin vector $\mathbf{OX}_2' = [\bar{1}01]_{hex\_t}$
- The parent vector **OY** is invariant, $\mathbf{OY} = [120]_{hex\_p}$ is equal to $\mathbf{OY}' = [120]_{hex\_t}$
- The parent vector $\mathbf{OG} = [\bar{1}01]_{hex\_p}$ is parallel to the twin vector $\mathbf{OG}' = [\bar{2}00]_{hex\_t}$

with the indices "p" and "t" in reference to the parent and twin bases. In their respective orthonormal bases these vectors are written

- $\mathbf{OX}_2 = [200]_{ortho\_p}$ and $\mathbf{OX'}_2 = [\bar{1}\,0\,\gamma]_{ortho\_t}$
- $\mathbf{OY} = [0\,\sqrt{3}\,0]_{ortho\_p}$ and $\mathbf{OY'} = [0\,\sqrt{3}\,0]_{ortho\_t}$
- $\mathbf{OG} = [\bar{1}01]_{hex\_p} = [\bar{1}\,0\,\gamma]_{ortho\_p}$ and $\mathbf{OG'} = [\bar{2}00]_{hex\_t} = [\bar{2}00]_{ortho\_t}$

In addition to the parallelism of the directions, the lengths of the vectors are such that $\|\mathbf{OX}_2\| \approx \|\mathbf{OX'}_2\|$, $\|\mathbf{OY}\| = \|\mathbf{OY'}\|$ and $\|\mathbf{OG}\| \approx \|\mathbf{OG'}\|$. Thus, the chosen vectors are relevant to build a model of twinning. We assume that the lattice distortion transforms the vectors as follows: $\mathbf{OX}_2 \to \mathbf{OX}_2'$, $\mathbf{OY} \to \mathbf{OY'}$, and $\mathbf{OG} \to \mathbf{OG'}$. The parent supercell $\mathbf{B}_{X_2YG} = (\mathbf{OX}_2, \mathbf{OY}, \mathbf{OG})$ is transformed into a distorted supercell $\mathbf{B}_{(X_2YG)'} = (\mathbf{OX'}_2, \mathbf{OY'}, \mathbf{OG'})$. The determinant of the matrix $(\mathbf{OX}_2, \mathbf{OY}, \mathbf{OG})$ gives ratio of the volume of the supercell by the volume of the unit cell (**a**, **b**, **c**). It is equal to 4. This can be compared with the volume ratios of the supercell XYZ used in the previous model of extension twinning [19] and that of the supercell XYE for contraction twinning [20], both equal to 2. The supercell XYG is thus twice larger than that used for extension and contraction twinning; which explains the difficulties we encountered when we tried to determine the trajectories of all the atoms of this supercell. Only hypothetic trajectories could be inferred for the moment; they are geometrically represented by the green curved arrows in Fig. 1b,c.

### 2.1. The coordinate transformation matrix

The coordinate transformation matrix is calculated by considering the rotation between the initial and final hexagonal lattices shown in Fig. 1b. The rotation angle is defined such that the image of the vector $\mathbf{OX}_2 = [200]_{ortho}$ becomes parallel to the vector $\mathbf{OG} = [\bar{1}\,0\,\gamma]_{ortho}$. This angle of rotation is thus $ArcCos(\frac{1}{\sqrt{1+\gamma^2}})$. It is 58.39° for magnesium or 57.78° for titanium. This rotation matrix is

$$\mathbf{R}^{p \to t}_{ortho} = \begin{pmatrix} \frac{1}{\sqrt{1+\gamma^2}} & 0 & \frac{\gamma}{\sqrt{1+\gamma^2}} \\ 0 & 1 & 0 \\ \frac{-\gamma}{\sqrt{1+\gamma^2}} & 0 & \frac{1}{\sqrt{1+\gamma^2}} \end{pmatrix} \tag{2}$$

### 2.2. The distortion matrix

The lattice distortion transforms the vectors $\mathbf{OX}_2 \to \mathbf{OX}_2'$, $\mathbf{OY} \to \mathbf{OY'}$, and $\mathbf{OG} \to \mathbf{OG'}$, as shown by straight green arrows in Fig. 1b. The supercells $\mathbf{B}_{X_2YG}$ and $\mathbf{B}_{(X_2YG)'}$ expressed in their local orthonormal basis $\mathbf{B}^p_{ortho}$ and $\mathbf{B}^t_{ortho}$ are

$$\mathbf{B}^p_{X_2YG} = \begin{pmatrix} 2 & 0 & -1 \\ 0 & \sqrt{3} & 0 \\ 0 & 0 & \gamma \end{pmatrix} \text{ and } \mathbf{B}^t_{(X_2YG)'} = \begin{pmatrix} -1 & 0 & 2 \\ 0 & \sqrt{3} & 0 \\ \gamma & 0 & 0 \end{pmatrix} \tag{3}$$

The supercell $\mathbf{B}_{(X_2YG)'}$ expressed in parent orthonormal basis is

$$\mathbf{B}^p_{(X_2YG)'} = \mathbf{T}^{p \to t}_{ortho} \mathbf{B}^t_{(X_2YG)'} = \begin{pmatrix} \sqrt{1+\gamma^2} & 0 & -\dfrac{2}{\sqrt{1+\gamma^2}} \\ 0 & \sqrt{3} & 0 \\ 0 & 0 & \dfrac{2\gamma}{\sqrt{1+\gamma^2}} \end{pmatrix} \quad (4)$$

Now that the vectors of these two supercells are expressed in the same basis $\mathbf{B}^p_{ortho}$, the distortion matrix $\mathbf{F}^{p \to t}_{ortho}$ written in the basis $\mathbf{B}^p_{ortho}$ is obtained by (see equation 1 of Ref. [18]):

$$\mathbf{F}^{p \to t}_{ortho} = \mathbf{B}^p_{(X_2YG)'} \cdot \left(\mathbf{B}^p_{X_2YG}\right)^{-1} \quad (5)$$

The calculations done with equations (3) and (4) lead to

$$\mathbf{F}^{p \to t}_{ortho} = \begin{pmatrix} \dfrac{\sqrt{1+\gamma^2}}{2} & 0 & \dfrac{\gamma^2 - 3}{2\gamma\sqrt{1+\gamma^2}} \\ 0 & 1 & 0 \\ 0 & 0 & \dfrac{2}{\sqrt{1+\gamma^2}} \end{pmatrix} \quad (6)$$

This active matrix can be expressed in the hexagonal basis $\mathbf{B}_{hex}$ by using the formula of coordinate change:

$$\mathbf{F}^{p \to t}_{hex} = \mathbf{H}^{-1}_{hex} \mathbf{F}^{p \to t}_{ortho} \mathbf{H}_{hex} = \begin{pmatrix} \dfrac{\sqrt{1+\gamma^2}}{2} & \dfrac{2-\sqrt{1+\gamma^2}}{4} & \dfrac{\gamma^2-3}{2\sqrt{1+\gamma^2}} \\ 0 & 1 & 0 \\ 0 & 0 & \dfrac{2}{\sqrt{1+\gamma^2}} \end{pmatrix} \quad (7)$$

This matrix is triangular; it has three eigenvalues: $\dfrac{\sqrt{1+\gamma^2}}{2}$, 1 and $\dfrac{2}{\sqrt{1+\gamma^2}}$, with three distinct eigenvectors that are $[100]_{hex} = \mathbf{OX}$, $[120]_{hex} = \mathbf{OY}$, and $[\bar{1}01]_{hex} = \mathbf{OG}$, respectively, as expected from Fig. 1. This is a stretch matrix in a non-orthogonal basis. Another way to reach the result could have been by noticing that

$$\frac{OX_2'}{OX_2} = \frac{\|[101]_{hex}\|}{\|[200]_{hex}\|} = \frac{\sqrt{1+\gamma^2}}{2} \quad (8)$$

$$\frac{OG'}{OG} = \frac{\|[\bar{2}00]_{hex}\|}{\|[\bar{1}01]_{hex}\|} = \frac{2}{\sqrt{1+\gamma^2}} \quad (9)$$

It can be checked that the determinant of the matrix (7) equals one, which is required to conserve the unit volume after distortion. In the case of an ideal ratio of $\gamma = \sqrt{8/3}$, the eigenvalues are

$$\mathbf{F}_{hex}^{p \to t} = \begin{pmatrix} \sqrt{\frac{11}{12}} & \frac{2 - \sqrt{\frac{11}{3}}}{4} & -\frac{1}{2\sqrt{33}} \\ 0 & 1 & 0 \\ 0 & 0 & \sqrt{\frac{12}{11}} \end{pmatrix} \approx \begin{pmatrix} 0.9574 & 0.0213 & -0.0870 \\ 0 & 1 & 0 \\ 0 & 0 & 1.0445 \end{pmatrix} \quad (10)$$

The values of the principal strains associated with this distortion matrix are then (-4.2%, 0, +4.4%). In the case of pure magnesium and pure titanium, they are (-4.6%, 0, +4.8%) and (-6.2%, 0, +6.6%), respectively. For a ratio $\gamma = \sqrt{3}$, the distortion matrix equals the identity matrix, i.e. there is no more strain at all because the contraction along **OX** and the extension along **OG** are both exactly compensated. For $\gamma > \sqrt{3}$, the signs of the strains become the opposite of those for $\gamma < \sqrt{3}$.

### 2.3. The correspondence matrix

The correspondence matrix $\mathbf{C}_{hex}^{t \to p}$ gives in the twin basis the distorted vectors of the parent basis. It is

$$\mathbf{C}_{hex}^{t \to p} = \mathbf{B}_{(X_2 YG)'}^{t} \cdot \left(\mathbf{B}_{X_2 YG}^{p}\right)^{-1} \quad (11)$$

The calculation is done by considering the vectors $\mathbf{OX_2}, \mathbf{OY}$, and $\mathbf{OG}$, and the vectors $\mathbf{OX_2}', \mathbf{OY}'$, and $\mathbf{OG}'$, in their respective hexagonal bases:

$$\mathbf{C}_{hex}^{t \to p} = \begin{pmatrix} 1 & 1 & -2 \\ 0 & 2 & 0 \\ 1 & 0 & 0 \end{pmatrix} \begin{pmatrix} 2 & 1 & -1 \\ 0 & 2 & 0 \\ 0 & 0 & 1 \end{pmatrix}^{-1} = \begin{pmatrix} \frac{1}{2} & \frac{1}{4} & -\frac{3}{2} \\ 0 & 1 & 0 \\ \frac{1}{2} & -\frac{1}{4} & \frac{1}{2} \end{pmatrix} \quad (12)$$

It can be checked that this matrix is unitary, i.e. it is equal to its inverse, as for other twinning correspondence matrices [7]. The correspondence matrix for the planes (reciprocal space) is directly deduced by the taking the inverse of the transpose of $\mathbf{C}_{hex}^{t \to p}$:

$$\left(\mathbf{C}_{hex}^{t \to p}\right)^* = \left(\mathbf{C}_{hex}^{t \to p}\right)^{-T} = \begin{pmatrix} \frac{1}{2} & 0 & -\frac{1}{2} \\ \frac{1}{4} & 1 & \frac{1}{4} \\ \frac{3}{2} & 0 & \frac{1}{2} \end{pmatrix} \quad (13)$$

The correspondence matrices (direct and reciprocal spaces) are independent of the ratio of lattice parameters $\gamma$, and is composed of rational values, as it should be.

The reader can verify the internal coherency of the calculations by checking the "master" equation

$$\mathbf{F}_{hex}^{p \to t} = \mathbf{T}_{hex}^{p \to t} \mathbf{C}_{hex}^{t \to p} \quad (14)$$

with $\mathbf{F}_{hex}^{p \to t}$ in equation (10), $\mathbf{C}_{hex}^{t \to p}$ in equation (12), and by writing $\mathbf{T}_{hex}^{p \to t} = \mathbf{H}_{hex}^{-1} \mathbf{T}_{ortho}^{p \to t} \mathbf{H}_{hex}$ with $\mathbf{T}_{ortho}^{p \to t}$ in equation (2) and $\mathbf{H}_{hex}$ in equation (1).

# 3. Convectional twins derived from the stretch twin by obliquity compensation

If a plane was let fully invariant by the distortion matrix $\mathbf{F}_{hex}^{p\rightarrow t}$, it would be a plane untilted by the distortion, i.e. it should be a eigenvector of the reciprocal distortion matrix. This matrix is

$$\left(\mathbf{F}_{hex}^{p\rightarrow t}\right)^* = \left(\mathbf{F}_{hex}^{p\rightarrow t}\right)^{-T} = \begin{pmatrix} \frac{2}{\sqrt{1+\gamma^2}} & 0 & 0 \\ \frac{1}{2} - \frac{1}{\sqrt{1+\gamma^2}} & 1 & 0 \\ -\frac{-3+\gamma^2}{2\sqrt{1+\gamma^2}} & 0 & \frac{\sqrt{1+\gamma^2}}{2} \end{pmatrix} \quad (15)$$

The eigenvalues are the same as for the direct distortion matrix. The associated eigenvectors are the planes $(001)_{hex}$, $(010)_{hex}$ and $(2\bar{1}2)_{hex}$. These are the only three planes that are not tilted. The $(001)_{hex}$ and $(2\bar{1}2)_{hex}$ planes are untilted but distorted because some of the directions they contain are rotated, or elongated or shorten. The prismatic plane $(010)_{hex}$ is the only invariant plane, but it has never been observed. Therefore, we have to introduce an additional rotation **R** with a low angle (called obliquity) that should combine with the distortion matrix in order to make a plane fully invariant (if the intermediate states are ignored). By using the correspondence matrix and the symmetry matrices of the hcp point group, it can be shown that only two planes can be fully restored; they are the planes $\boldsymbol{g_a} = (\bar{2}12)_{hex}$ and $\boldsymbol{g_b} = (2\bar{1}6)_{hex}$. Therefore, the obliquity made by their tilt during the distortion $\mathbf{D}_{hex}^{p\rightarrow t}$ should be compensated. The details of the calculations are given in the two next sections.

## 3.1. Obliquity compensation required for the conventional $(\bar{2}12)$ twin mode

The obliquity is given by a rotation that has for axis **OY** and its angle should compensate the rotation of the plane $(\bar{2}12)_{hex}$. This angle noted $\xi_a$ is the angle made by the reciprocal vector $\boldsymbol{g_a} = (\bar{2}12)_{hex}$ with its image $\boldsymbol{g'_a}$ by the distortion. The calculations are done in the orthonormal basis. The angle $\xi_a$ is thus the angle between $\mathbf{H}_{hex}^* \boldsymbol{g_a}$ and $\left(\mathbf{F}_{ortho}^{p\rightarrow t}\right)^* . \mathbf{H}_{hex}^* \boldsymbol{g_a}$, with $\left(\mathbf{F}_{ortho}^{p\rightarrow t}\right)^*$ the inverse of the transpose of the matrix (6). The calculations (see *SupplData1*) shows that

$$\xi_a = ArcCos\left(\frac{-1+3\gamma^2}{(1+\gamma^2)^{3/2}}\right) \quad (16)$$

For a hard-sphere packing ratio $\gamma = \sqrt{\frac{8}{3}}$, the obliquity is $\xi_a = ArcCos\left(\frac{21}{11}\sqrt{\frac{3}{11}}\right) \approx 4.45°$. It is 4.82° and 6.64° for pure magnesium and titanium, respectively.

The rotation matrix of axis $\mathbf{OY} = [010]_{ortho}$ and angle $-\xi_a$ expressed in $\mathbf{B}_{ortho}$ is

$$\mathbf{R}^a = \begin{pmatrix} \frac{-1+3\gamma^2}{(1+\gamma^2)^{3/2}} & 0 & -\frac{\gamma(-3+\gamma^2)}{(1+\gamma^2)^{3/2}} \\ 0 & 1 & 0 \\ \frac{\gamma(-3+\gamma^2)}{(1+\gamma^2)^{3/2}} & 0 & \frac{-1+3\gamma^2}{(1+\gamma^2)^{3/2}} \end{pmatrix} \quad (17)$$

The distortion matrix corrected of this obliquity is

$$\mathbf{D}^a_{ortho} = \mathbf{R}^a \, \mathbf{F}^{p \to t}_{ortho} = \begin{pmatrix} \frac{3}{2} - \frac{2}{1+\gamma^2} & 0 & -\frac{-3+\gamma^2}{2(\gamma+\gamma^3)} \\ 0 & 1 & 0 \\ \gamma(\frac{1}{2} - \frac{2}{1+\gamma^2}) & 0 & \frac{1}{2} + \frac{2}{1+\gamma^2} \end{pmatrix} \quad (18)$$

In the hexagonal basis this matrix becomes

$$\mathbf{D}^a_{hex} = \mathbf{H}^{-1}_{hex} \, \mathbf{D}^a_{ortho} \, \mathbf{H}_{hex} = \begin{pmatrix} \frac{3}{2} - \frac{2}{1+\gamma^2} & -\frac{1}{4} + \frac{1}{1+\gamma^2} & -\frac{1}{2} + \frac{2}{1+\gamma^2} \\ 0 & 1 & 0 \\ \frac{1}{2} - \frac{2}{1+\gamma^2} & -\frac{1}{4} + \frac{1}{1+\gamma^2} & \frac{1}{2} + \frac{2}{1+\gamma^2} \end{pmatrix} \quad (19)$$

For a hard-sphere packing ratio $\gamma = \sqrt{\frac{8}{3}}$ this matrix takes the values

$$\mathbf{D}^a_{hex} = \begin{pmatrix} \frac{21}{22} & \frac{1}{44} & \frac{1}{22} \\ 0 & 1 & 0 \\ -\frac{1}{22} & \frac{1}{44} & \frac{23}{22} \end{pmatrix} \approx \begin{pmatrix} 0.9545 & 0.0227 & 0.0455 \\ 0. & 1. & 0. \\ -0.0455 & 0.0227 & 1.0455 \end{pmatrix} \quad (20)$$

The new orientation of the twin is

$$\mathbf{R}^{p \to t}_{ortho} \, \mathbf{R}^a = \begin{pmatrix} 1 - \frac{2}{1+\gamma^2} & 0 & \frac{2\gamma}{1+\gamma^2} \\ 0 & 1 & 0 \\ -\frac{2\gamma}{1+\gamma^2} & 0 & 1 - \frac{2}{1+\gamma^2} \end{pmatrix} \quad (21)$$

Which is a rotation of axis **OY** = **a**+2**b** and angle $ArcCos\left(1 - \frac{2}{1+\gamma^2}\right)$. For a hard-sphere packing ratio $\gamma = \sqrt{\frac{8}{3}}$, this angle takes the value $ArcCos\left(\frac{5}{11}\right) \approx 62.96°$. The angle becomes is 63.21° for magnesium, and 64.43° for titanium. This proves that the prototype (58°, **a**+2**b**) twinning model corrected from its obliquity becomes the classical $\{11\bar{2}2\}$ twinning mode reported frequently in titanium and zirconium alloy. As pointed in the introduction, a similar observation was made by Ostapovets *et al*. [24] by using a double-shear model.

The shear vector and amplitude can be calculated by applying the matrix $\mathbf{D}^a_{ortho}$ of equation (18) to the unit vector normal to the $(\bar{2}12)_{hex}$ plane, i.e. $\mathbf{n}^a = \frac{1}{\sqrt{1+\gamma^2}}[-\gamma, 0, 1]_{ortho}$. The calculations show that

$$\mathbf{s}^a = (\mathbf{D}^a_{ortho} - \mathbf{I}).\mathbf{n}^a = \frac{(3-\gamma^2)}{2\gamma\sqrt{1+\gamma^2}}[1, 0, \gamma]_{ortho} \quad (22)$$

The shear vector is thus parallel to $[1\,0\,1]_{hex}$ and the shear amplitude is

$$s^a = \|\mathbf{s}^a\| = \frac{(3-\gamma^2)}{2\gamma} \tag{23}$$

For a hard-sphere packing ratio $\gamma = \sqrt{\frac{8}{3}}$, this shear takes the value $s^a \approx 0.102$. The shear is $s^a = \frac{9}{4\sqrt{583}} \approx 0.111$ for magnesium, and 0.152 for titanium. These results are compared to the WR model of $\{11\bar{2}2\}$ twin in section 4.

### 3.2. Obliquity compensation required for the conventional $(2\bar{1}6)$ twin mode

The obliquity is given by a rotation that has for axis $\mathbf{OY}$ and its angle should compensate the rotation of the plane $(2,\bar{1},6)_{hex}$. This angle noted $\xi_b$ is the angle made by the reciprocal vector $\boldsymbol{g_b} = (2,\bar{1},6)_{hex}$ with its image $\boldsymbol{g'_b}$ by the distortion. The angle $\xi_a$ is thus the angle between $\mathbf{H}^*_{hex}\boldsymbol{g_b}$ and $\left(\mathbf{F}^{p\rightarrow t}_{ortho}\right)^*.\mathbf{H}^*_{hex}\,\boldsymbol{g_b}$, with $\left(\mathbf{F}^{p\rightarrow t}_{ortho}\right)^*$ the inverse of the transpose of the matrix (6). The calculations (see **SupplData1**) shows that

$$\xi_b = ArcCos\left(\frac{9+5\gamma^2}{\sqrt{1+\gamma^2}(9+\gamma^2)}\right) \tag{24}$$

For a hard-sphere packing ratio $\gamma = \sqrt{\frac{8}{3}}$, the obliquity is $\xi_b = ArcCos\left(\frac{67}{35}\sqrt{\frac{3}{11}}\right) \approx 1.39°$. It is 1.51°° and 2.03° for pure magnesium and titanium, respectively.

The rotation matrix of axis $\mathbf{OY} = [010]_{ortho}$ and angle $-\xi_b$ expressed in $\mathbf{B}_{ortho}$ is

$$\mathbf{R}^b = \begin{pmatrix} \frac{9+5\gamma^2}{\sqrt{1+\gamma^2}(9+\gamma^2)} & 0 & \frac{\gamma(-3+\gamma^2)}{\sqrt{1+\gamma^2}(9+\gamma^2)} \\ 0 & 1 & 0 \\ -\frac{\gamma(-3+\gamma^2)}{\sqrt{1+\gamma^2}(9+\gamma^2)} & 0 & \frac{9+5\gamma^2}{\sqrt{1+\gamma^2}(9+\gamma^2)} \end{pmatrix} \tag{25}$$

The distortion matrix corrected of the obliquity is

$$\mathbf{D}^b_{ortho} = \mathbf{R}^b\,\mathbf{F}^{p\rightarrow t}_{ortho} = \begin{pmatrix} \frac{5}{2}-\frac{18}{9+\gamma^2} & 0 & -\frac{3}{2\gamma}+\frac{6\gamma}{9+\gamma^2} \\ 0 & 1 & 0 \\ \gamma(-\frac{1}{2}+\frac{6}{9+\gamma^2}) & 0 & -\frac{1}{2}+\frac{18}{9+\gamma^2} \end{pmatrix} \tag{26}$$

In the hexagonal basis this matrix becomes

$$\mathbf{D}^b_{hex} = \mathbf{H}^{-1}_{hex}\,\mathbf{D}^b_{ortho}\,\mathbf{H}_{hex} = \begin{pmatrix} \frac{5}{2}-\frac{18}{9+\gamma^2} & -\frac{3}{4}+\frac{9}{9+\gamma^2} & \frac{9}{2}-\frac{54}{9+\gamma^2} \\ 0 & 1 & 0 \\ -\frac{1}{2}+\frac{6}{9+\gamma^2} & \frac{1}{4}-\frac{3}{9+\gamma^2} & -\frac{1}{2}+\frac{18}{9+\gamma^2} \end{pmatrix} \tag{27}$$

For a hard-sphere packing ratio $\gamma = \sqrt{\frac{8}{3}}$ this matrix takes the values

$$\mathbf{D}^b_{hex} = \begin{pmatrix} \frac{67}{70} & \frac{3}{140} & -\frac{9}{70} \\ 0 & 1 & 0 \\ \frac{1}{70} & -\frac{1}{140} & \frac{73}{70} \end{pmatrix} \approx \begin{pmatrix} 0.9571 & 0.0214 & -0.1286 \\ 0 & 1 & 0 \\ 0.0143 & -0.0071 & 1.0429 \end{pmatrix} \quad (28)$$

The new orientation of the twin is

$$\mathbf{R}^{p \to t}_{ortho} \mathbf{R}^b = \begin{pmatrix} -1 + \frac{18}{9+\gamma^2} & 0 & \frac{6\gamma}{9+\gamma^2} \\ 0 & 1 & 0 \\ -\frac{6\gamma}{9+\gamma^2} & 0 & -1 + \frac{18}{9+\gamma^2} \end{pmatrix} \quad (29)$$

Which is a rotation of axis **OY** = **a**+2**b** and angle $ArcCos\left(-1 + \frac{18}{9+\gamma^2}\right)$. For a hard-sphere packing ratio $\gamma = \sqrt{\frac{8}{3}}$, this angle takes the value $ArcCos\left(\frac{19}{35}\right) \approx 57.12°$. The angle becomes is 56.88° for magnesium, and 55.75° for titanium. To our knowledge, this twinning mode is not reported in the observed studies on Ti and Zr alloys. For magnesium alloys, some histograms of misorientations extracted from EBSD maps reported in literature exhibit a peak corresponding to rotation of angle close to 56° and axis close to **a**+2**b** [22][23], and Ostapovets *et al.* [24] observed the corresponding twins by EBSD and TEM.

The shear vector and amplitude can be calculated by applying the matrix $\mathbf{D}^b_{ortho}$ of equation (18) to the unit vector normal to the $(2,\bar{1},6)_{hex}$ plane, i.e. $\mathbf{n}^b = \frac{1}{\sqrt{9+\gamma^2}}[\gamma, 0, 3]_{ortho}$. The calculations show that

$$\mathbf{s}^b = (\mathbf{D}^b_{ortho} - \mathbf{I}) \cdot \mathbf{n}^b = \frac{(3-\gamma^2)}{2\gamma\sqrt{9+\gamma^2}} [\bar{3}, 0, \gamma]_{ortho} \quad (30)$$

The shear vector is thus parallel to $[-3\ 0\ 1]_{hex}$ and the shear amplitude is

$$s^b = \|\mathbf{s}^b\| = \frac{(3-\gamma^2)}{2\gamma} \quad (31)$$

The shear value is exactly the same for the $\{11\bar{2}2\}$ twins, as expected for shear that differs only by their obliquity correction.

## 4. Discussions

The model of $\{11\bar{2}2\}$ and $\{11\bar{2}6\}$ twinning modes presented in this paper is based on a (58°,**a**+2**b**) prototype stretch twin; the two modes are conjugate and the difference between them only come from the value of the obliquity correction. This approach is the same as for the previous models of extension and contraction twinning [19][20]. The supercell chosen for the calculations are twice that used for these modes, which introduce more complexity and impeded us for the moment to find the analytical equations of the atomic trajectories. Thus, contrarily to Ref. [19][20], only the distortion associated with the complete distortion process could be calculated; the continuous analytical expression of the distortion matrix, the maximum volume change during the distortion, the prediction on the formation of twins remain to determine. Despite this limitation, interesting information can already be extracted from the model. The calculated twin/parent misorientations related to the $\{11\bar{2}2\}$ and $\{11\bar{2}6\}$ twins are rotations around **a+2b** axis; the rotation angles are in the case of $\{11\bar{2}2\}$ twinning 63.21° for magnesium and 64.43° for titanium, and in the case of $\{11\bar{2}6\}$ twinning 56.88° for magnesium and 55.75° for titanium. The shear values for both $\{11\bar{2}2\}$ and $\{11\bar{2}6\}$ twins are the same, i.e. 0.102 for hard-sphere packing, 0.111 for magnesium and 0.152 for titanium. These results are the same as those recently obtained by Ostapovets *et al.* [24], despite the difference of the initial assumption in the mechanism. Their model assumes that the twins result from a $\{10\bar{1}2\} - \{10\bar{1}2\}$ double-twin process, and that primary extension twin was completely consumed by the second extension twin because there is no trace of it in the EBSD maps. As the composition of two shears is in general not a shear but an invariant line strain, they had to introduce an additional rotation in order to get an invariant plane (simple shear) plane. This additional rotation is similar to the obliquity compensation used here. The difference between the two models thus lies in the initial distortion; it is a stretch in our model and a double-shear in Ostapovets *et al.* 's model. The advantage of the stretch representation is that it is possible to graphically understand the distortion and get an idea of the atom trajectories, even if the analytical calculations are yet too complex to be analytically calculated. The other point is that the approach is the same as that of extension and contraction twinning; the distortion can be obtained in one step. We hope to prove in a next future that the so-called $\{10\bar{1}1\} - \{10\bar{1}2\}$ double-twins are also the result of a one-step process.

Whatever the mechanism, we agree with Ostapovets *et al.* to say that the $\{11\bar{2}2\}$ and $\{11\bar{2}6\}$ twinning modes are "*can be considered as new twinning modes in magnesium*". They are not predicted by the classical theory of twinning [5]-[9]; the shear values are the lowest values reported for hcp metals for the moment (we will see that twins with lower values in Ref. [25]); they are even lower than that of extension twinning. Indeed, whatever the packing ratio, the shear value associated with $\{11\bar{2}2\}$ and $\{11\bar{2}6\}$ twinning is $s = \frac{(3-\gamma^2)}{2\gamma}$, whereas that extension twinning is $s = \frac{(3-\gamma^2)}{\sqrt{3}\gamma}$; the ratio between the two shear modes $\frac{2}{\sqrt{3}} \approx 1.15$, indepedntly of the $\gamma$ value. One can wonder why such twins were not predicted despite the fact that they are conventional, i.e. expressed by a simple shear matrix. The reason is not yet clear to us. The size of the supercell chosen in our model does not seem to be the reason because larger supercells were considered by Bevis and Crocker [7]. We propose the possible explanation. There is a weak point in Bevis and Crocker's calculations; it is the way the symmetries are treated. Instead of considering all the equivalent correspondence matrices by using the twenty four symmetry matrices in the point group of the hcp phase, they proceeded by *"interchanging rows, interchanging columns, changing the signs of rows and changing the signs of columns",* which clearly forgets many other possibilities. Further work would be required to check whether or not the new $\{11\bar{2}2\}$ and $\{11\bar{2}6\}$ twinning modes could be predicted by the Bevis and Crocker's theory after modifying the way with which symmetries are treated.

Our model of $\{11\bar{2}2\}$ twinning can also be compared with the earlier WR model. The WR model assumes that $\{11\bar{2}2\}$ twinning is realized by the coordinated displacements of dislocations in a periodicity of three $\{11\bar{2}2\}$ planes. If one compare our model with WR model in Fig. 3 and tries to express WR model in term of supercell, it can be noticed that the main difference rely on the choice of the supercell. The supercell associated with our model is slightly larger than theirs because the shear in our model is done on the fourth $\{11\bar{2}2\}$ plane, and not on the third one. Increasing the supercell size implies more shuffle, but the gain here is important as the shear value ($s \approx 0.111$) is twice lower than that of obtained by the WR model ($s \approx 0.219$). This could explain why in titanium alloys the $\{11\bar{2}2\}$ twins can be more frequent than $\{10\bar{1}2\}$ extension twins ($s \approx 0.174$), which is difficult to understand with WR model. As the shear directions are opposite between the two models (compare the direction of the coloured arrows in Fig. 3a and Fig. 3b), one can imagine some experiments on a single crystal to confront the models. We predict that $\{11\bar{2}2\}$ twins are extension twins in magnesium or titanium, whereas WR model predicts that they are contraction twins. For the moment, it can be noticed EBSD observations showed that $\{10\bar{1}2\}$ twins and $\{11\bar{2}2\}$ twins can coexist in the same grains in Ti alloys [2], and in the same areas in magnesium single crystal [25]; which is a good hint that $\{11\bar{2}2\}$ twins are extension twins.

The model is thus promising, but some questions remain to answer. Why $\{11\bar{2}2\}$ twins are rarely observed in magnesium polycrystalline alloys in comparison with extension twins? Why $\{11\bar{2}2\}$ twins are more frequent than $\{11\bar{2}6\}$ twins in titanium alloys if their obliquity angle is larger and if their shear amplitudes are the same? The response to the former question is probably linked to the highest complexity and highest number of shuffling atoms of the supercell XYG required for $\{11\bar{2}2\}$ twinning. It can also be expected that the volume change involved for $\{11\bar{2}2\}$ twinning is higher than for $\{10\bar{1}2\}$ extension twinning; but that point will be clearer only once the atomic trajectories will be determined. The response to the latter question is probably to be found in the way the material can accommodate the distortion imposed by the twinning process. This opens the way for future researches to explore dislocation plasticity in hcp metals induced by twinning distortion.

The present paper has presented two classical (shear) modes derived from the (58°, **a**+2**b**) prototype stretch model. A third mode, observed in a magnesium single crystal, will be shown to be unconventional (not shear); the observations and the theoretical calculations related to this mode will be presented in a separate paper [25].

## 5. Conclusion

This paper proposes a crystallographic model for the $\{11\bar{2}2\}$ and $\{11\bar{2}6\}$ twinning modes. These modes were recently put in evidence by Ostapovets *et al.* (Phil. Mag, 2017) and interpreted as the result of $\{10\bar{1}2\} - \{10\bar{1}2\}$ double-twinning mechanism with simultaneous action of two twinning shears. Here, they are interpreted as a one-step mechanism based on a 58°, **a**+2**b**) prototype stretch twin. The two twin modes differ from the prototype only by their slight obliquity correction. This study also brings complementary information to Ostapovets *et al.* 's paper. The twins are explained geometrically; the misorientation matrices, the correspondence matrices, and the distortion matrices with their corresponding shear values are calculated analytically as function of the packing ratio. The results are compared with classical theory of twinning and with Westlake-Rosenbaum's model of $\{11\bar{2}2\}$ twinning. An unconventional twinning mode recently discovered in a magnesium single crystal and based on the same prototype stretch twin will be the subject of a separate publication.

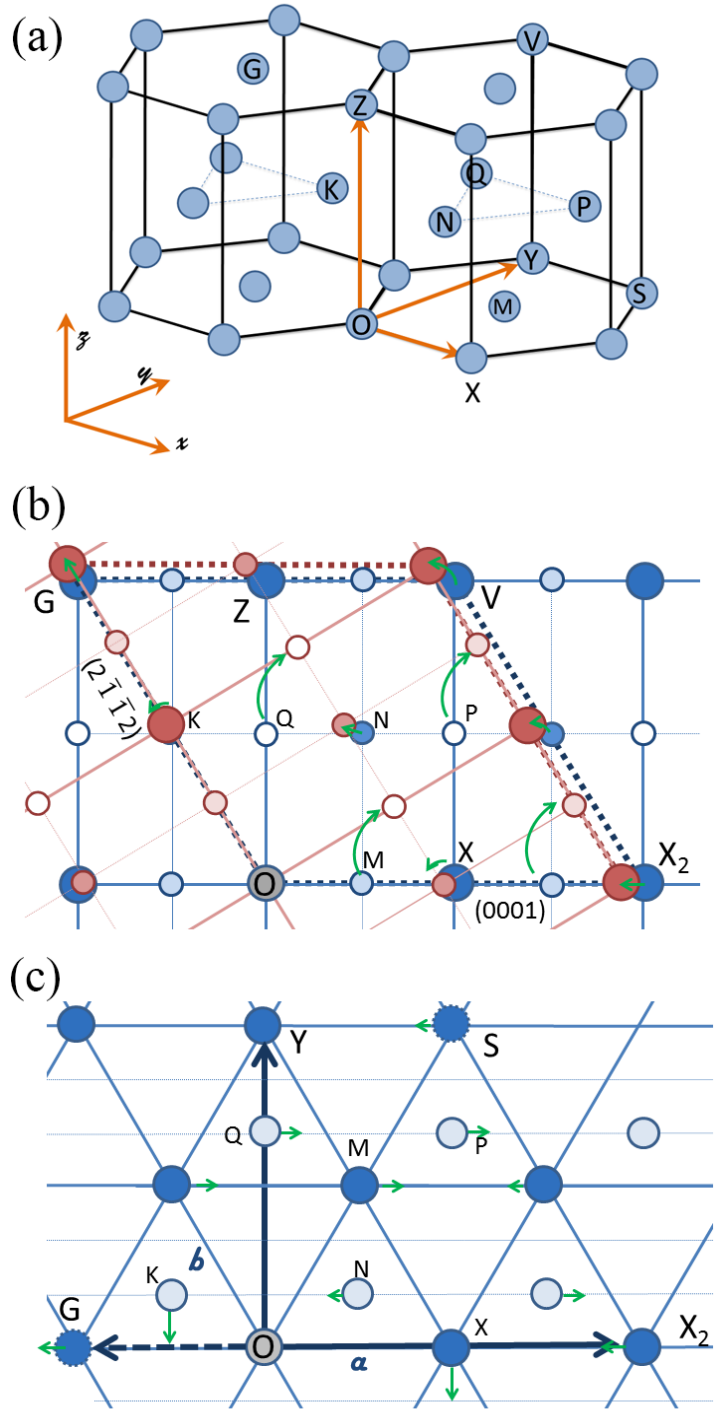

Fig. 1. Schematic view of the (58°, a+2b) prototype stretch twin. (a) 3D figure with the xyz orthogonal basis formed on the XYZ cell. The supercell used for the calculation is XYG. Twinning distortion viewed in projection: (b) along the **OY** = [120]$_{hex}$ axis, and (c) along the **OZ** = [001]$_{hex}$ axis. The possible atomic displacements are indicated by the green straight or curved arrows. The distortion of the supercell XYG is marked by the green straight arrows.

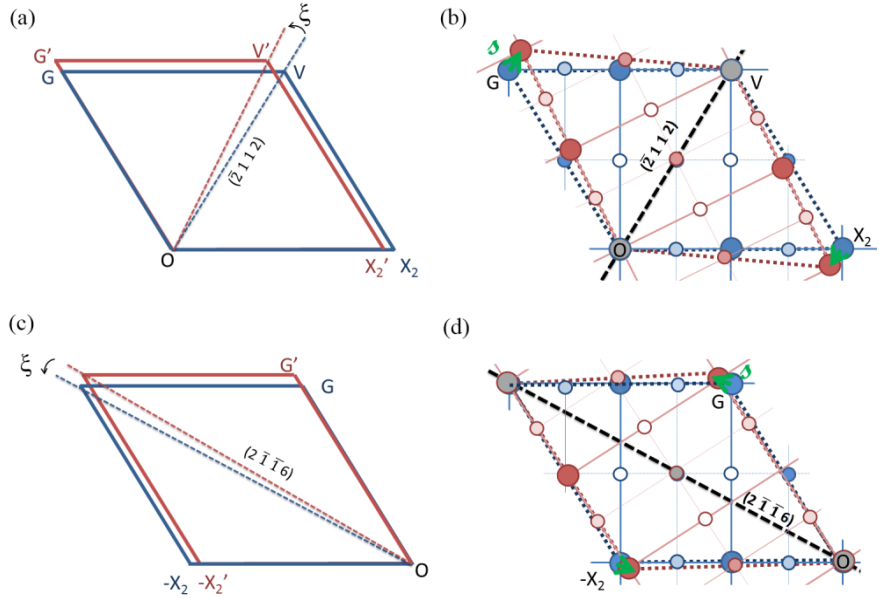

Fig. 2. The two possible obliquity corrections required to transform the (58°, ***a+2b***) stretch twin into a simple shear twin, by maintaining untilted (a, b) the plane ($\bar{2}112$), (c, d) plane ($2\bar{1}\bar{1}6$). The obliquity angle $\xi$ before correction is shown in (a, c), and after correction in (b, d).

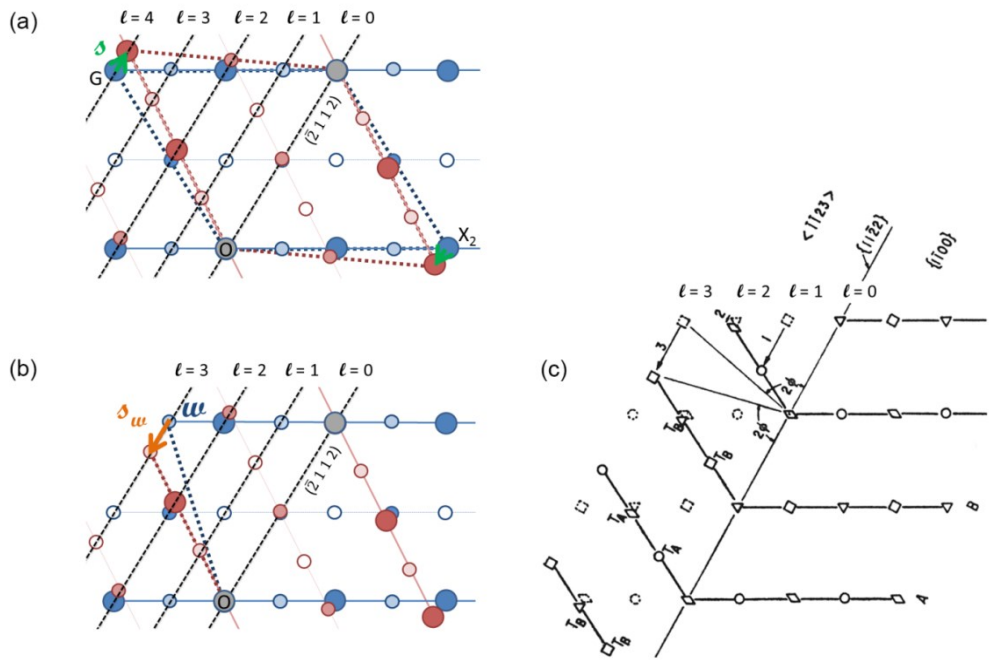

Fig. 3. Comparison with the WR model of $\{11\bar{2}2\}$ twinning. (a) Shear vector corresponding to the present model. (b) Shear vector obtained with the WR model [10][11]. (c) Original scheme shown in Ref. [11]. The three « levels » of the $\{11\bar{2}2\}$ plane are indicated to help the identification between (b) and (c).